\documentclass[conference,10pt]{IEEEtran}
\IEEEoverridecommandlockouts
\usepackage{cite}
\usepackage{subfigure}
\usepackage{amsmath,amssymb,amsfonts}
\usepackage{caption}
\usepackage{subfigure}
\usepackage{algorithmic}
\usepackage[ruled,linesnumbered]{algorithm2e}
\usepackage{multirow}
\usepackage{graphicx}
\usepackage{textcomp}
\usepackage{float}
\usepackage{color}
\usepackage{array}
\usepackage{ifpdf}
\def\BibTeX{{\rm B\kern-.05em{\sc i\kern-.025em b}\kern-.08em
    T\kern-.1667em\lower.7ex\hbox{E}\kern-.125emX}}
\begin{document}

\title{Energy Efficiency Optimization in IRS-Enhanced mmWave Systems with Lens Antenna Array\\
}

\author{\IEEEauthorblockN{Yazheng Wang$^*$, Hancheng Lu$^*$, Dan Zhao$^*$, Huan Sun$^\dagger$}
\IEEEauthorblockA{$^*$Key Laboratory of Wireless-Optical Communications, University of Science and Technology of China, Hefei, China. \\ $^\dagger$Wireless Technology Laboratory, Huawei Technologies Co., Ltd., Shanghai, China. \\
Email: $^*$WANG1997@mail.ustc.edu.cn, $^*$hclu@ustc.edu.cn, $^*$zd2019@mail.ustc.edu.cn, $^\dagger$sunhuan11@huawei.com}
}

\maketitle

\begin{abstract}\label{abstract}
In millimeter wave (mmWave) systems, the advanced lens antenna array can effectively reduce the radio frequency chains cost. However, the mmWave signal is still vulnerable to blocking obstacles and suffers from severe path loss. To address this problem, we propose an intelligent reflect surface (IRS) enhanced multi-user mmWave communication system with lens antenna array. Moreover, we attempt to optimize energy efficiency in the proposed system. An energy efficiency maximization problem is formulated where the transmit beamforming at base station and the reflect beamforming at IRSs are jointly considered. To solve this non-convex problem, we propose an algorithm based on the alternating optimization technique. In the proposed algorithm, the transmit beamforming is handled by the sequential convex approximation method and the reflect beamforming is optimized based on the quadratic transform method. Our simulation results show that the proposed algorithm can achieve significant energy efficiency improvement under various scenarios.
\end{abstract}


\vspace{-0.5em}
\section{Introduction}\label{introduction}
The fifth-generation system is supposed to serve abundant devices and simultaneously achieve the 1000x capacity increase at a similar or lower power consumption compared with current wireless communication system \cite{surveyofee}. Millimeter wave (mmWave) communication (\emph{i.e.}, the frequency ranges from 30 to 300GHz) is considered as a complementary method to meet this increasing demand in capacity due to the large available spectrum\cite{mmw}. However, high frequency inevitably causes strong pathloss and a large number of antennas are required to deploy at the base station (BS) to compensate the severe signal attenuation. Fortunately, the small wavelengths of mmWave signal makes it possible to arrange massive antennas in a limited physical space to achieve high beamforming gain. Such a large antenna array system leads to unaffordably high hardware and energy cost since the traditional multiple input multiple output (MIMO) requires one radio frequency (RF) chain for each antenna. To overcome this problem, an advanced technique to utilize is \emph{lens antenna array}\cite{beamlens,lens, MUlens, eelens, widebandlens}. 

Lens antenna array is generally composed of an electromagnetic lens and a vast number of antennas located in the focal region of the lens. According to \cite{beamlens}, lens antenna array is used to effectively reduce the RF chains cost by transforming the signal from the antenna space to the beam space with lower dimensions. In \cite{lens}, the array response of the lens antenna array is proved to be a ``sinc" function which leads to the angle-dependent energy focusing property, \emph{i.e.}, the peak response of the array is determined by the angle of departure (AoD) or the angle of arrival (AoA) of the signal. An uplink mmWave single-sided lens MIMO system is analysed in \cite{MUlens} and a low-complexity path division multiple access scheme is proposed. Furthermore, the authors in \cite{widebandlens} maximize the minimum signal-to-interference-plus-noise ratio (SINR) with limited RF chains via antenna selection and beamforming design.

In spite of the benefits brought by lens antenna array, mmWave systems are still susceptible to blocking obstacles where the line-of-sight (LOS) links can be easily broken. \emph{Intelligent reflect surface} (IRS) \cite{irscover}, as a new concept beyond massive MIMO, is a promising hardware technology to create additional LOS links and enhance the mmWave lens system. Specifically, an IRS consists of a large number of nearly passive reflecting elements, which are capable of independently changing the phase shifts of the incident signal with low power consumption. By smartly adjusting the elements, the signal reflected by IRSs can be added coherently at the receiver to improve the received signal power. The researches on the IRS-enhanced communication systems have recently captured a lot attention\cite{irscover,sumrate,irsrsrp,eeirs,reflectarray,lischannel}. The authors in \cite{sumrate} study the weighted sum rate maximization problem and design three low-complexity algorithms for the reflect beamforming at IRS. In \cite{eeirs}, an energy efficiency (EE) problem is optimized by exploiting zero-forcing (ZF) precoding and power allocation. Furthermore, a practical hardware platform for mmWave is designed in \cite{reflectarray}.

In this paper, we propose an IRSs-enhanced multi-user (MU) mmWave lens system by integrating IRS with lens antenna array, with the goal to exploit benefits from these two technologies simultaneously. To improve the communication quality without incurring high energy consumption, we attempt to optimize the EE of our proposed system. We consider an MU downlink mmWave scenario with lens, and multiple IRSs are deployed to assist the system. In addition, instead of continuous phase shifters, we adopt IRSs with discrete phase shifters which are more practical. Moreover, an alternating optimization based algorithm is exploited to optimize the transmit beamforming at BS and the reflect beamforming at IRSs. Therefore, the main contributions of this paper can be summarized as follows:
\begin{itemize}
\item We propose an IRSs-enhanced MU mmWave lens system. Particularly, lens antenna array is utilized at BS to reduce the RF chain numbers and multiple IRSs with discrete phase shifters are deployed to create reliable LOS links without incurring huge energy consumption. 
\item We formulate an EE maximization problem by taking into account the constraints of maximum data transmission power, individual quality of service (QoS) and limited RF chains number.
\item Due to the non-convexity and complexity of the problem, we develop an alternating optimization based algorithm to solve it. Specifically, the reflect beamforming at IRSs is optimized by utilizing the quadratic transform method and the transmit beamforming is handled through the sequential convex approximation (SCA) method.
\end{itemize}

Extensive simulations are carried out to validate our work, and the simulation results show that our proposed algorithm can achieve significant EE enhancement compared with other schemes under various scenarios. 

The rest of this paper is organized as follows. In Section \uppercase\expandafter{\romannumeral2} we describe related system models and formulate an EE maximization problem. In Section \uppercase\expandafter{\romannumeral3}, an algorithm based on the alternating optimization technique is exploited to solve the problem. Simulation results are conducted in Section \uppercase\expandafter{\romannumeral4}. Finally, in Section \uppercase\expandafter{\romannumeral5} we conclude this paper.

\section{System Model}

\begin{figure}[!t]
	\centering
	\includegraphics[width=.9\columnwidth,height=6.5cm]{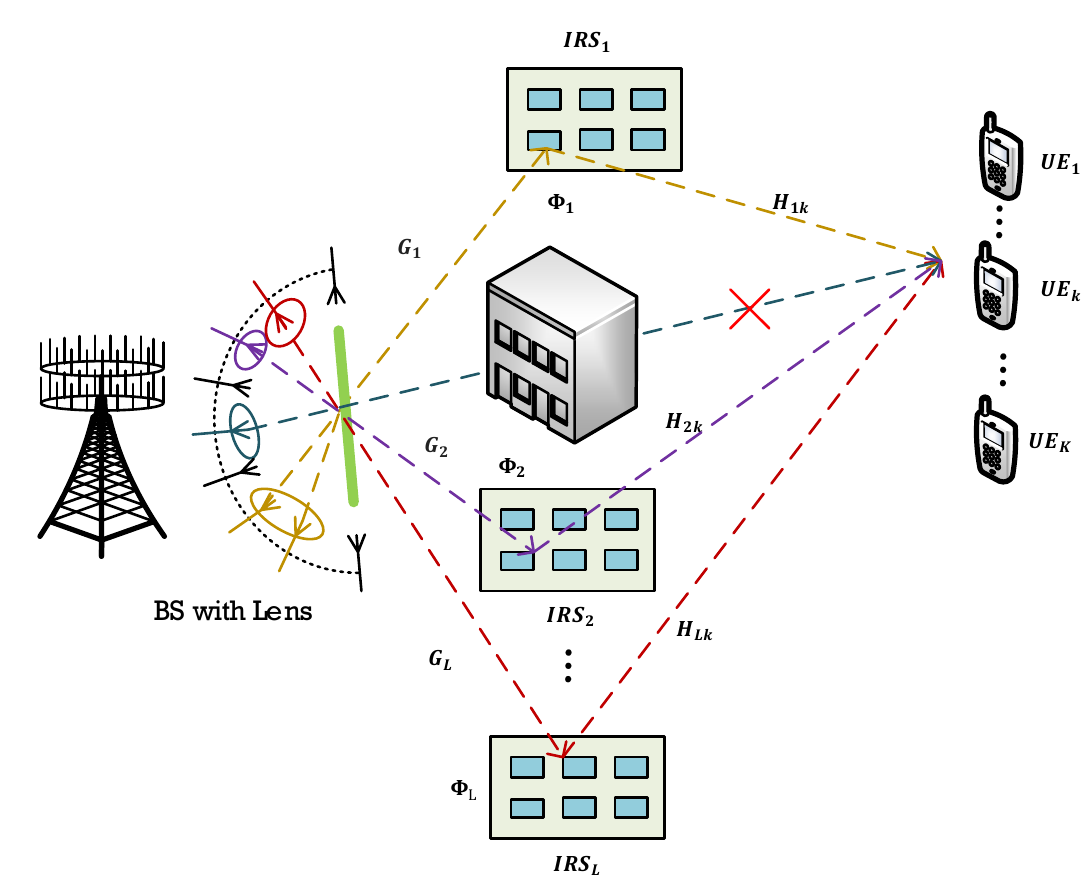}
	\vspace{-0.5em}
	\caption{IRSs-enhanced MU mmWave lens system.}
	\label{fig:framework}
	\vspace{-1.5em}
\end{figure}

As shown in Fig. \ref{fig:framework}, we consider an IRSs-enhanced MU mmWave lens system, where the lens at BS is equipped with $M$ antennas. A total of $L$ IRSs are deployed to serve $K$ single-antenna users and each IRS consists of $N$ reflect elements. We further define $\mathcal{K} \triangleq \{1,2,\cdots,K\}$ and $\mathcal{L} \triangleq \{1,2,\cdots,L\}$ as the set of users and IRSs. The direct links between BS and users are supposed to be severely blocked by obstacles.

\subsection{IRS-assisted Transmission}
The received signal of the $k$-th user can be expressed as
\begin{equation}
	\label{recv signal}
	\begin{split}
		y_k&=\sum_{l\in \mathcal{L}}\mathbf{H}_{lk}^{H} \boldsymbol{\Phi}_l^H \mathbf{G}_l\boldsymbol{\omega}_k s_k+ \\ &\sum_{l\in \mathcal{L}}\mathbf{H}_{lk}^{H} \boldsymbol{\Phi}_l^H \mathbf{G}_l \sum_{i\in \mathcal{K},i\neq k}\boldsymbol{\omega}_i s_i +n_k,	
		\vspace{-0.7em}
	\end{split}
\end{equation}
where $\mathbf{G}_l \in \mathbb{C}^{N\times M}$ is the channel matrix between BS and the $l$-th IRS, and $\mathbf{H}_{lk}^H \in \mathbb{C}^{1\times N}$ is the channel matrix between the $l$-th IRS and the $k$-th user, the matrix of the $l$-th IRS is denoted by $\boldsymbol{\Phi}_l =\rm diag\{\Gamma e^{j\theta_{l1}},\Gamma e^{j\theta_{l2}},\cdots,\Gamma e^{j\theta_{lN}}\}$ and $\theta_{ln}$ represents the phase shift associated with the $n$-th passive reflect element on the $l$-th IRS. For simplicity, the amplitude reflection coefficient $\Gamma$ is set as 1 in the rest of this paper. The parameter $n_k \sim \mathcal{CN}(0,\sigma^2)$ denotes the complex additive white Gaussian noise with zero mean and variance $\sigma^2$, and $s_k$ is independent data symbol with zero mean and unit power, while the beamforming vector for the $k$-th user at BS is denoted by $\boldsymbol{\omega}_k \in \mathbb{C}^{M\times 1}$. The power for data transmission is subject to maximum transmit power constraint
\begin{equation}
	\label{powerst}
	\begin{split}
		\sum_{k\in \mathcal{K}} \left|\left|\boldsymbol{\omega}_k\right|\right|^2_{2} \leqslant P_T,
		\vspace{-1em}
	\end{split}
\end{equation}
where $\left|\left|*\right|\right|_{2}$ represents the Euclidean norm and $P_T$ is the maximum transmit power at BS. We further define the lens beamforming matrix $\mathbf{W} \in \mathbb{C}^{M\times K}$ as
\vspace{-0.5em}
\begin{equation} \label{beammatrixdefine}
	\boldsymbol{W} = \left[{\boldsymbol{\omega}_{1}},{\boldsymbol{\omega}_{2}},\cdots,{\boldsymbol{\omega}_{K}}\right]=
	\left[\begin{matrix}
	\underbrace{{{w}_{11}},{{w}_{12}},\cdots,{{w}_{1K}}}_{{{\boldsymbol{\tilde{\omega}}}_{1}}} \\
	\vdots  \\
	\underbrace{{{w}_{M1}},{{w}_{M2}},\cdots,{{w}_{MK}}}_{{{\boldsymbol{\tilde{\omega}}}_{M}}} \\
	\end{matrix} \right],
\end{equation}
where $\boldsymbol{\tilde{\omega}}_m$ denotes the beamforming vector on the $m$-th antenna. To express the constraint of limited RF chains at BS, we define
\vspace{-0.5em}
\begin{equation} \label{beamdefine}
\bar{\boldsymbol{\omega}}={{\left[ {{\left\| {{{\boldsymbol{\tilde{\omega}}}}_{1}} \right\|}_{\infty }},{{\left\| {{{\boldsymbol{\tilde{\omega}}}}_{2}} \right\|}_{\infty }},...,{{\left\| {{{\boldsymbol{\tilde{\omega}}}}_{M}} \right\|}_{\infty }} \right]}^{H}},
\end{equation}
where $\left|\left|*\right|\right|_{\infty}$ represents the $l_{\infty}$-norm. Specifically, if the $m$-th antenna is not selected for transmission (\emph{i.e.}, it is not connected to any RF chain), then the beamforming vector on the $m$-th antenna is forced to be zero (\emph{i.e.}, ${{\boldsymbol{\tilde{\omega}}}}_{m} = \mathbf{0}$ ). Due to the limited RF chains constraint at BS, we should have $\left\| \bar{\boldsymbol{\omega}} \right\|_0 \leqslant N_{RF}$, where $N_{RF}$ is the RF chains number and $\left|\left|*\right|\right|_{0}$ is the $l_0$-norm.

It follows from (\ref{recv signal}) that the SINR for the $k$-th user can be expressed as
\vspace{-1em}
\begin{equation}\label{SINRk}
\gamma_k=\frac{\left|(\sum_{l\in \mathcal{L}}\mathbf{H}_{lk}^{H} \boldsymbol{\Phi}_l^H \mathbf{G}_l)\boldsymbol{\omega}_k\right|^2}{\left|(\sum_{l\in \mathcal{L}}\mathbf{H}_{lk}^{H} \boldsymbol{\Phi}_l^H \mathbf{G}_l)\sum_{i\in \mathcal{K},i\neq k}\boldsymbol{\omega}_i\right|^2+\sigma^2}.
\end{equation} 
then the sum rate of all users is given by
\vspace{-0.1em}
\begin{equation}\label{sumrate}
	R_{sum}=W\sum_{k\in \mathcal{K}}\log\left(1+\gamma_k\right),
	\vspace{-0.3em}
\end{equation}
where $W$ is the bandwidth.
\subsection{Power Consumption Model}
We adopt the power models presented in \cite{eelens} and \cite{eeirs}. Therefore, the total power consumption can be described as
\vspace{-0.2em}
\begin{equation} \label{powermodel}
	P_{total}=\epsilon\sum_{k\in \mathcal{K}} \left\|\boldsymbol{\omega}_k\right\|^2_{2}+N_{RF}P_{RF}+LNP_N(B)+P_{cir},
	\vspace{-0.5em}
\end{equation}
where $\epsilon=\frac{1}{\xi}$, $\xi \in (0,1]$ is the power amplifier efficiency. $P_{RF}$ is the power consumed by each RF chain. $P_N(B)$ is the power dissipated by each reflect element on each IRS for $B$-bit phase resolution, and $LNP_N(B)$ is the total power consumption of all IRSs. The total circuits power consumption is denoted by $P_{cir}$. We define $P_c \triangleq N_{RF}P_{RF}+LNP_N(B)+P_{cir}$ in the rest of this paper. 


\subsection{Lens Channel Model}
Since the direct links between BS and users are severely blocked by obstacles, only the channels between BS and IRSs, the channels between IRSs and users are taken into account. We model the BS-IRS channels according to the following geometric channel model
\vspace{-0.5em}
\begin{equation}\label{channelmodel}
	\mathbf{G}_l=\sum_{g=0}^{G_{p}}K_g\mathbf{a}_{l,IRS}(\phi_{lg,AoA})\mathbf{a}_{LAA}^H(\phi_{lg,AoD}),
	\vspace{-0.5em}
\end{equation}
where $K_g$ is the channel gain and $g=0$ denotes the LOS path, $G_p$ is the number of NLOS paths. A uniform linear array (ULA) is deployed at IRSs\cite{lischannel}. The parameters $\phi_{lg,AoA}$ and $\phi_{lg,AoD}$ represent the AoA and the AoD of the signal reflected by the $l$-th IRS in the $g$-th path. The unique array response of lens antenna array $\mathbf{a}_{LAA}$ can be expressed according to \cite{lens}
\begin{equation}\label{lens}
	a_{m}(\phi_{l,AoD})=\sqrt{A} {\rm sinc}(m-\tilde{D}\sin\phi_{l,AoD}), m \in \mathcal{M},
\end{equation}
where $A$ is the normalized apertures and $\tilde{D}$ is the lens's normalized dimension\cite{lens}, while $\mathcal{M}\triangleq \{0,\pm1,\cdots,\pm\frac{M-1}{2}\}$ denotes the set of antenna indices.

For the channel between the $k$-th user and the $l$-th IRS, we assume that the power of LOS path is much higher than the sum of the power of NLOS paths \cite{irscover}, and the channel can be modeled as
\begin{equation}\label{channelmodel2}
\mathbf{H}_{lk}=\beta_{lk} \mathbf{a}_{l,IRS}(\phi_{lk,AoD}),
\end{equation}
where $\beta_{lk}$ indicates the channel gain and $\phi_{lk,AoD}$ is the AoD of the signal from the $l$-th IRS to the $k$-th user.

\subsection{Problem Formulation}\label{problemformat}
We aim to maximize the EE of the proposed system which is defined as the ratio of the sum rate and the total power consumption. With the models presented above, we can formulate the EE maximization problem as
\begin{subequations}\label{maxee}
	\begin{align}
	\underset{\left\{{\boldsymbol{\omega}_{k}} \right\}_{k\in \mathcal{K}},\{\boldsymbol{\Phi}_l\}_{l\in \mathcal{L}}}{\mathop{\max}}\, ~~& \frac{W\sum_{k\in \mathcal{K}}\log\left(1+\gamma_k\right)}{\epsilon\sum_{k\in \mathcal{K}}\left\|\boldsymbol{\omega}_k\right\|^2_{2}+P_c}, \label{ee}\\
	s.t.~~& \sum_{k\in \mathcal{K}} \left|\left|\boldsymbol{\omega}_k\right|\right|^2_{2} \leqslant P_T, \label{ptst}\\[-0.5mm]
	& \theta_{ln} \in \mathcal{F},\label{irsst}\\[-0.5mm]
	& \gamma_k \geqslant \rho_k , \forall k \in \mathcal{K}, \label{QoS} \\
	& \left\| \bar{\boldsymbol{\omega}} \right\|_0 \leqslant N_{RF}, \label{RFst}
	\end{align}
\end{subequations}
where $ \mathcal{F}=\left\{ 0,\frac{2\pi }{{{2}^{B}}},\frac{2\pi *2}{{{2}^{B}}},\cdots,\frac{2\pi *({{2}^{B}}-1)}{{{2}^{B}}} \right\}$ is the set of available phase shifts for each IRS element and $B$ is the resolution of the phase shifter. Due to the non-convex objective function (\ref{ee}) and the non-convex constraints (\ref{irsst}), (\ref{RFst}), this problem is generally more complicated than the problems formulated in \cite{sumrate,irscover,eeirs,irsrsrp}. In the sequel, we try to solve this problem efficiently.


\section{Energy Efficiency Optimization}
In this section, we propose an algorithm to handle the formulated EE maximization problem based on alternating optimization technique.
\vspace{-0.5em}
\subsection{Reflect Beamforming Optimization}
In this subsection, we investigate how to optimize IRS matrices $\{\boldsymbol{\Phi}_l\}_{l\in \mathcal{L}}$ with given transmit beamforming. When $\{\boldsymbol{\omega}_k\}_{k\in \mathcal{K}}$ are fixed, the denominator of (\ref{ee}) becomes a constant. Dropping the irrelevant constant $W$, the problem can be transformed to a sum rate maximization problem
\vspace{-0.2em}
\begin{equation} \label{maxsumrate}
	\begin{split}
	\underset{\{\boldsymbol{\Phi}_l\}_{l\in \mathcal{L}}}{\mathop{\max}}\, ~~& f(\{\boldsymbol{\Phi}_l\}_{l\in \mathcal{L}})=\sum_{k\in \mathcal{K}}\log\left(1+\gamma_k\right), \\[-1.5mm]
	s.t.~~& \theta_{ln} \in \mathcal{F}, \\
	& \gamma_k \geqslant \rho_k, \forall k \in \mathcal{K}.
	\end{split}
	\vspace{-0.5em}
\end{equation}

By introducing auxiliary variables $\boldsymbol{\lambda}=\left[\lambda_1,\lambda_2,\cdots,\lambda_k\right]^T$, we equivalently represent the objective function as
\vspace{-0.2em}
\begin{equation}\label{f1}
	f_1(\{\boldsymbol{\Phi}_l\}_{l\in \mathcal{L}}, \boldsymbol{\lambda})=\sum_{k\in \mathcal{K}}\log(1+\lambda_k)-\sum_{k\in \mathcal{K}}\lambda_k+\sum_{k\in \mathcal{K}}\frac{(1+\lambda_k)\gamma_k}{(1+\gamma_k)}.
\end{equation}

When $\{\boldsymbol{\Phi}_l\}_{l\in \mathcal{L}}$ are fixed, the optimal $\lambda_k$ is $\lambda_k^{opt}=\gamma_k$. Then for a fix $\boldsymbol{\lambda}$, we only need to cope with the last term in the right side of (\ref{f1}). Combining the expression of $\gamma_k$ in (\ref{SINRk}), the objective function can be expressed as
\vspace{-0.5em}
\begin{equation}\label{f2}
	f_2(\{\boldsymbol{\Phi}_l\}_{l\in \mathcal{L}})=\sum_{k\in \mathcal{K}} \frac{(1+\lambda_k)\left|(\sum_{l\in \mathcal{L}}\mathbf{H}_{lk}^{H}\boldsymbol{\Phi}_l^{H}\mathbf{G}_l)\boldsymbol{\omega}_k\right|^2}{\sum_{i\in \mathcal{K}}\left|(\sum_{l\in \mathcal{L}}\mathbf{H}_{lk}^{H}\boldsymbol{\Phi}_l^{H}\mathbf{G}_l)\boldsymbol{\omega}_i\right|^2+\sigma^2}.
\end{equation}

Since $\boldsymbol{\Phi}_l$ is a diagonal matrix, we observe that $\mathbf{H}_{lk}^{H}\boldsymbol{\Phi}_l^{H}\mathbf{G}_l\boldsymbol{\omega}_k=\boldsymbol{\varphi}_l^{H}{\rm diag}(\mathbf{H}_{lk}^{H})\mathbf{G}_l\boldsymbol{\omega}_k$, where
$\boldsymbol{\varphi}_l^H=[e^{-j\theta_{l1}},e^{-j\theta_{l2}},\cdots,e^{-j\theta_{lN}}]$. Define $\boldsymbol{a}_{k,i,l}={\rm diag}(\mathbf{H}_{lk}^H)\mathbf{G}_l\boldsymbol{\omega}_i$,
$\mathbf{v}_{k,i}^H=[\mathbf{a}_{k,i,1}^H,\mathbf{a}_{k,i,2}^H,\cdots,\mathbf{a}_{k,i,L}^H]$, and
$\mathbf{u}^H=[\boldsymbol{\varphi}_1^H,\boldsymbol{\varphi}_2^H,\cdots,\boldsymbol{\varphi}_{L}^H]$. 
Then the problem (\ref{maxsumrate}) can be transformed to the following problem
\vspace{-0.5em}
\begin{equation}\label{fixrho2}
\begin{split}
\underset{\mathbf{u}}{\mathop{\max}}\, ~~& f_3(\mathbf{u})=\sum_{k\in \mathcal{K}}\frac{(1+\lambda_k)\left|\mathbf{u}^H\mathbf{v}_{k,k}\right|^2}{\sum_{i\in \mathcal{K}}\left|\mathbf{u}^H\mathbf{v}_{k,i}\right|^2+\sigma^2}, \\[-0.5mm]
s.t.~~& \theta_{ln} \in \mathcal{F},\\
& \gamma_k \geqslant \rho_k, \forall k \in \mathcal{K}.
\end{split}
\end{equation}

We notice that this is a multiple-ratio fractional programming problem which can be solved by the quadratic transform method \cite{fp}. Therefore, we reformulate the objective function as follows
\vspace{-0.5em}
\begin{equation}\label{f4}
\begin{split}
f_4(\mathbf{u})&=\sum_{k\in \mathcal{K}}2\sqrt{1+\lambda_k}\text{Re}\{y_k^*\mathbf{u}^H\mathbf{v}_{k,k}\}- \\ &\sum_{k\in \mathcal{K}}\left|y_k\right|^2(\sum_{i\in \mathcal{K}}\left|\mathbf{u}^H\mathbf{v}_{k,i}\right|^2+\sigma^2),
\end{split}
\end{equation}
and $\mathbf{y}=[y_1,y_2,\cdots,y_k]^{T}$ is an auxiliary variable vector. Then we alternatively optimize $\mathbf{y}$ and $\mathbf{u}$. With given $\mathbf{u}$, the optimal $y_k$ can be obtained by setting $\partial f_4/\partial {y_k}$ to zero
\vspace{-0.3em}
\begin{equation}
	y_k^{opt}=\frac{\sqrt{1+\lambda_k}(\mathbf{u}^H\mathbf{v}_{k,k})}{\sum_{i\in \mathcal{K}}\left|\mathbf{u}^H\mathbf{v}_{k,i}\right|^2+\sigma^2}.
\end{equation}
\vspace{-0.5em}

By utilizing the relationship that $\left|\mathbf{u}^H\mathbf{v}_{k,i}\right|^2=\mathbf{u}^H\mathbf{v}_{k,i}\mathbf{v}^H_{k,i}\mathbf{u}$, the problem can be rewritten as
\vspace{-0.3em}
\begin{subequations}\label{fixy}
\begin{align}
\underset{\mathbf{u}}{\mathop{\max}}\, ~~& f_5\left(\mathbf{u}\right)=-\mathbf{u}^H\mathbf{A}\mathbf{u}+2\text{Re}\{\mathbf{u}^H\mathbf{B}\}+\mathbf{C}, \\
s.t.~~& \theta_{ln} \in \mathcal{F},\\
& \mathbf{u}^{H}\mathbf{D}\mathbf{u}-\rho_k\sigma^2 \geqslant 0 , \forall k \in \mathcal{K}, \label{QoS3}
\vspace{-0.2em}
\end{align}
\end{subequations}
where
\begin{equation}
\label{AB}
\begin{split}
&\mathbf{A}=\sum_{k\in \mathcal{K}}\left|y_k\right|^2\sum_{i\in \mathcal{K}}\mathbf{v}_{k,i}\mathbf{v}_{k,i}^H , 
\quad \mathbf{B}=\sum_{k\in \mathcal{K}}(\sqrt{1+\lambda_k})y^*_k\mathbf{v}_{k,k} , \\
&\mathbf{C}=-\sum_{k\in \mathcal{K}}\left|y_k\right|^2\sigma^2 ,\quad\quad \mathbf{D}=\mathbf{v}_{k,k}\mathbf{v}_{k,k}^{H}-\sum_{i\in \mathcal{K},i\neq k}\rho_k\mathbf{v}_{k,i}\mathbf{v}_{k,i}^{H}.
\vspace{-2em}
\end{split}
\end{equation}
Similar to \cite{sumrate}, we iteratively optimize one of the elements of $\mathbf{u}$ by keeping the other $NL-1$ elements fixed. Since $\mathbf{A}$ is a Hermitian matrix, we can obtain
\vspace{-0.5em}
\begin{equation}
\label{nthphase}
\begin{aligned}[b]
& \mathbf{u}^H\mathbf{A}\mathbf{u}=u_n^*A_{n,n}u_n+\sum_{i=1,i \neq n}^{NL}\sum_{j=1,j \neq n}^{NL}u_i^*A_{i,j}u_j\\
&\qquad+2\operatorname{Re}\{\sum_{j=1,j \neq n}^{NL}u_n^*A_{n,j}u_j\} , \\
& \mathbf{u}^H\mathbf{B}=u_n^*B_n+\sum_{i=1,i \neq n}^{NL}u_i^*B_i ,
\vspace{-0.5em}
\end{aligned}
\end{equation}
where $u_n$ denotes the $n$-th element of $\mathbf{u}$ and $u_n^{*}u_n=1$. The element at the $i$-th row and the $j$-th column of $\mathbf{A}$ and the $i$-th element of $\mathbf{B}$ are denoted by $A_{i,j}$ and $B_i$. We substitute (\ref{nthphase}) into (\ref{fixy}) and drop the irrelevant constant, the problem (\ref{fixy}) can be transformed to
\vspace{-1.0em}
\begin{subequations}
\begin{align}
\underset{\angle u_n}{\mathop{\max}}\, ~~& -A_{n,n}+2\text{Re}\{u_n^*(B_n-\sum_{j=1,j \neq n}^{NL}A_{n,j}u_j)\},\\[-2mm]
s.t.~~& \angle u_n \in \mathcal{F},\\
& \mathbf{u}^{H}\mathbf{D}\mathbf{u}-\rho_k\sigma^2 \geqslant 0 , \forall k \in \mathcal{K}. \label{DQoS}
\vspace{-1.5em}
\end{align}
\end{subequations}

Define $d_n \triangleq B_n-\sum_{j=1,j \neq n}^{NL}A_{n,j}u_j$ and use $\angle d_n$ to represent the argument of $d_n$. The argument of $u_n$ is denoted by $\angle u_n$, which also indicates the phase shift of the IRS reflect element. The optimal solution of $u_n$ can be obtained through
\vspace{-0.5em}
\begin{equation}\label{opttheta}
\angle u_{n}^{opt}=\arg \underset{\angle u_n \in \mathcal{F}_{QoS}}{\mathop{\min }}\,\left| \angle d_n -\angle u_n \right|.
\vspace{-0.5em}
\end{equation}

To obtain the feasible set $\mathcal{F}_{QoS}$, we select all the phase shifts which satisfy the constraint (\ref{DQoS}) from the finite set $\mathcal{F}$.
\vspace{-1.5em}
\subsection{Transmit Beamforming Optimization}
Given fixed IRSs matrices $\{\boldsymbol{\Phi}_l\}_{l\in \mathcal{L}}$, we attempt to optimize $\{\boldsymbol{\omega}_k\}_{k\in \mathcal{K}}$ with the constraints of maximum data transmission power, individual QoS and limited RF chains. A fully digital beamforming for IRSs-enhanced MU system with $N_{RF}=M$ is first obtained by SCA\cite{optimalee}. Then based on the obtained fully digital beamforming results, the \emph{Power Based Antenna Selection} \cite{widebandlens} is utilized to effectively accommodate the limited RF chains, without incurring severe performance degradation.

For the special case with full RF chains at BS, the constraint (\ref{RFst}) is guaranteed to be satisfied and hence can be removed. 
By dropping the irrelevant constant variable $W$, we rewrite the problem as
\vspace{-0.5em}
 \begin{subequations}\label{fixirs}
 	\begin{align}
 	\underset{\left\{{\boldsymbol{\omega}_{k}} \right\}_{k\in \mathcal{K}}}{\mathop{\max}}\, ~~& \frac{\sum_{k\in \mathcal{K}}\log\left(1+\gamma_k\right)}{\epsilon\sum_{k\in \mathcal{K}} \left\|\boldsymbol{\omega}_k\right\|^2_{2}+P_c}, \\
 	s.t.~~& \sum_{k\in \mathcal{K}} \left|\left|\boldsymbol{\omega}_k\right|\right|^2_{2} \leqslant P_T, \\
 	& \gamma_k \geqslant \rho_k, \forall k \in \mathcal{K} \label{qos}
 	\end{align}
  \end{subequations}

Then we introduce variables $\eta$ and $t$ denoting the squared EE and the squared power consumption, respectively. Thus the problem (\ref{fixirs}) can be further equivalently transformed as
\vspace{-0.5em}
\begin{subequations}\label{fixirs1}
	\begin{align}
	{\mathop{\max}}\, ~~& \eta \\
	s.t.~~& \sum_{k\in \mathcal{K}}\log(1+\gamma_k) \geqslant \sqrt{\eta t}, \label{nconv}\\[-1.0mm]
	& \sqrt{t} \geqslant \epsilon\sum_{k\in \mathcal{K}} \left\|\boldsymbol{\omega}_k\right\|^2_{2}+P_c, \label{tst}\\[-1.0mm]
	& \sum_{k\in \mathcal{K}} \left|\left|\boldsymbol{\omega}_k\right|\right|^2_{2} \leqslant P_T, \label{Pst} \\[-0.7mm]
	& \frac{1}{\sqrt{\rho_k}}\text{Re}\{\mathbf{H}_k^{H}\boldsymbol{\omega}_k\} \geqslant (\sum_{i\in \mathcal{K},i\neq k}\left|\mathbf{H}_k^{H}\boldsymbol{\omega}_i\right|^2+\sigma^2)^{\frac{1}{2}} , \forall k \in \mathcal{K}. \label{QoS2}\
	\vspace{-0.2em}
	\end{align}
\end{subequations}
where $\mathbf{H}_k^{H}=\sum_{l\in \mathcal{L}}\mathbf{H}_{lk}^{H}\boldsymbol{\Phi}_l^{H}\mathbf{G}_l$ denotes the combined channel for the $k$-th user and the constraint (\ref{qos}) is transformed to (\ref{QoS2}). Besides, without loss of optimality, we assume the optimal solution $\boldsymbol{\omega}_k^{opt}$ makes $\mathbf{H}_{k}^{H}\boldsymbol{\omega}_k$ a real number for all $k$\cite{widebandlens}. It can be observed that constraints (\ref{tst}), (\ref{Pst}) and (\ref{QoS2}) are convex. To tackle the non-convex constraint (\ref{nconv}), we transform it as follows
\vspace{-0.2em}
\begin{subequations}\label{nconv1}
\begin{align}
& \sum_{k\in \mathcal{K}} a_k \geqslant \sqrt{\eta t}, \label{nconv11}\\
& \log(v_k) \geqslant a_k, \forall k \in \mathcal{K}, \label{nconv12}\\
& 1+\gamma_k \geqslant v_k, \forall k \in \mathcal{K}, \label{nconv13}
\vspace{-0.7em}
\end{align}
\end{subequations}
where $\mathbf{a}=\left[a_1,a_2,\cdots,a_K\right]^{T}$ and $\mathbf{v}=\left[v_1,v_2,\cdots,v_K\right]^{T}$ are new variables and $a_k$ is the rate of the $k$-th user. Furthermore, the constraint (\ref{nconv13}) can be expressed as
\begin{subequations}\label{nconv13_1}
\begin{align}
& \text{Re}\{\mathbf{H}_{k}^{H}\boldsymbol{\omega}_k\} \geqslant \sqrt{(v_k-1)b_k}, \forall k \in \mathcal{K}, \label{nconv13_11} \\
& b_k \geqslant \sigma^2+\sum_{i\in \mathcal{K},i\neq k}\left|\mathbf{H}_{k}^{H}\boldsymbol{\omega}_i\right|^2, \forall k \in \mathcal{K}, \label{nconv13_12}
\vspace{-0.4cm}
\end{align}
\end{subequations}
where $\mathbf{b}=\left[b_1,b_2,\cdots,b_K\right]^{T}$ are new variables which can represent the inter-user interference plus noise. 
With optimization variables $\{\boldsymbol{\omega}_k\}_{k\in \mathcal{K}}$, $\eta$, $t$, $\mathbf{a}$, $\mathbf{v}$, $\mathbf{b}$, the non-convexity comes from constraints (\ref{nconv11}) and (\ref{nconv13_11}) which have the similar form and can be solved by SCA. We denote the value of variable $z$ after the $m$-th step of iteration by $z^{(m)}$. Take the constraint (\ref{nconv11}) as an example, the function $\sqrt{\eta t}$ is obviously jointly concave with respect to $\eta$ and $t$ for all $\eta \geqslant 0$ and $t \geqslant 0$. Thus the right side of the constraint (\ref{nconv11}) can be replaced by a convex upper approximation
\vspace{-0.3em}
\begin{equation}\label{fa}
	\sqrt{\eta t} \leqslant \sqrt{\eta^{(m)} t^{(m)}}+\frac{1}{2}\sqrt{\frac{t^{(m)}}{\eta^{(m)}}}(\eta-\eta^{(m)})+\frac{1}{2}\sqrt{\frac{\eta^{(m)}}{t^{(m)}}}(t-t^{(m)}).
\end{equation}

Let $f_a(\eta^{(m)},t^{(m)})$ denotes the right side of (\ref{fa}) and it actually represents the first order of the function $\sqrt{\eta t}$ around point $(\eta^{(m)},t^{(m)})$. In our formulation, the first order approximation in (\ref{fa}) is well defined over the whole feasible set of (\ref{fixirs1}) since variables $\eta$ and $t$ are strictly positive. Similar to (\ref{fa}), we utilize $f_a(v_k^{(m)},b_k^{(m)})$ to replace the right side of constraint (\ref{nconv13_11}). Then the problem (\ref{fixirs1}) can be approximated at the $m$-th step of iteration by the following convex problem
\vspace{-0.4em}
\begin{subequations}\label{fixirs2_mstep}
	\begin{align}
	{\mathop{\max}}\, ~~& \eta \\
	s.t.~~& \text{Re}\{\mathbf{H}_{k}^{H}\boldsymbol{\omega}_k\} \geqslant f_a(v_k^{(m)},b_k^{(m)}) , \forall k \in \mathcal{K}, \\
	& \sum_{k\in \mathcal{K}} a_k \geqslant f_a(\eta^{(m)},t^{(m)}), \\
	& (\ref{tst}),(\ref{Pst}),(\ref{QoS2}),(\ref{nconv12}),(\ref{nconv13_12}).
	\end{align}
\end{subequations}

Problem (\ref{fixirs2_mstep}) is a generalized nonlinear convex problem, which can be tackled by available solvers such as Fmincon. With given IRS matrices $\{\boldsymbol{\Phi}_l\}_{l\in \mathcal{L}}$, the problem (\ref{fixirs}) is optimized via Algorithm \ref{scaalgorithm}, \emph{i.e.}, we iteratively solve the convex problem (\ref{fixirs2_mstep}) to approximate the solution. 
\begin{algorithm}
	\caption{SCA beamforming with given IRS matrices}\label{scaalgorithm}
	Initialization: Feasible set $(\eta^{(0)}$,$t^{(0)}$,$\mathbf{v}^{(0)}$,$\mathbf{b}^{(0)})$ and set m=0. \\
	\Repeat{convergence}{
		Solve problem (\ref{fixirs2_mstep}) with $(\eta^{(m)}$,$t^{(m)}$,$\mathbf{v}^{(m)}$,$\mathbf{b}^{(m)})$ and obtain the optimal solution $(\eta^{opt}$,$t^{opt}$,$\mathbf{v}^{opt}$,$\mathbf{b}^{opt})$.\\
		m :=m+1. \\
		Update $\eta^{(m)}=\eta^{opt}$,$t^{(m)}=t^{opt}$,$\mathbf{v}^{(m)}=\mathbf{v}^{opt}$,$\mathbf{b}^{(m)}=\mathbf{b}^{opt}$
		
	}
	\KwOut{The optimal beamforming vectors $\{\boldsymbol{\omega}_k^{opt}\}_{k=1}^K$.}
\end{algorithm}
\vspace{-0.5em}

To cope with the non-convex sparse constraint $\left\| \bar{\boldsymbol{\omega}} \right\|_0 \leqslant N_{RF}$, the optimal fully digital beamforming vector $\{\boldsymbol{\omega}_k^{opt}\}_{k=1}^K$ is first obtained via Algorithm \ref{scaalgorithm}. Then we select those $N_{RF}$ antennas contributing most to the power, \emph{i.e.}, $N_{RF}$ antennas with the most significant transmit power $P_m$. Specially, $P_m$ can be calculated by $P_m=\sum_{k=1}^{K}\left|\omega_{mk}^{opt}\right|^2$ where $\omega_{mk}^{opt}$ is the $m$-th element of $\boldsymbol{\omega}_k^{opt}$. After selecting the most significant $N_{RF}$ antennas, the optimal solution corresponding to those $N_{RF}$ antennas can be similarly obtained via Algorithm \ref{scaalgorithm}.
\vspace{-0.2em}

\section{Simulation Results}
We assume the IRSs-enhanced MU mmWave lens system operates at 28GHz with bandwidth $W=500$ MHz. The BS equipped with lens is located at the origin point and users are randomly distributed in a circle at $(100 m,0 m)$ with radius $30m$. In addition, the total number of antennas at lens is $M = 151$. We consider $L = 2$ IRSs which are located at $(50 m, 50 m)$ and $(50 m, -50m)$, respectively. For the power consumption model, we set $\epsilon=1.2$, $P_{RF}=300$mW and $P_{cir}=200$mW. Typical power consumption for each IRS element are $6, 7.8$mW for $5$-, and $6$-bit phase resolution\cite{eeirs}. The channel gain $\beta_{lk}$ is taken as $\beta_{lk}=\alpha_l g_t g_r$ where $g_t=9.82$dBi, $g_r=0$dBi represent the transmit and receive antenna gain, $\alpha_l \sim \mathcal{CN}(0,10^{-0.1\kappa})$ and $\kappa=\kappa_a+10\kappa_b\log(d)+\kappa_c$ with $\kappa_a=61.4, \kappa_b=2, \kappa_c \sim \mathcal{N}(0,\sigma^2_c)$ and $\sigma_c=5.8$dB. The LOS path gain $G_0$ is set the same as $\beta_{lk}$, while the values of NLOS path gain are set as $\kappa_{a^{'}}=72, \kappa_{b^{'}}=2.92, \sigma_{c^{'}}=8.7$dB. Besides, we set $P_T=30$dbm, $\sigma^2=-117$dbm, $B=6$ and $\rho_k=0$dB, $\forall k \in \mathcal{K}$ as default values.

\begin{figure}
	[t]
	\centering
	\includegraphics[width=.9\columnwidth,height=6cm]{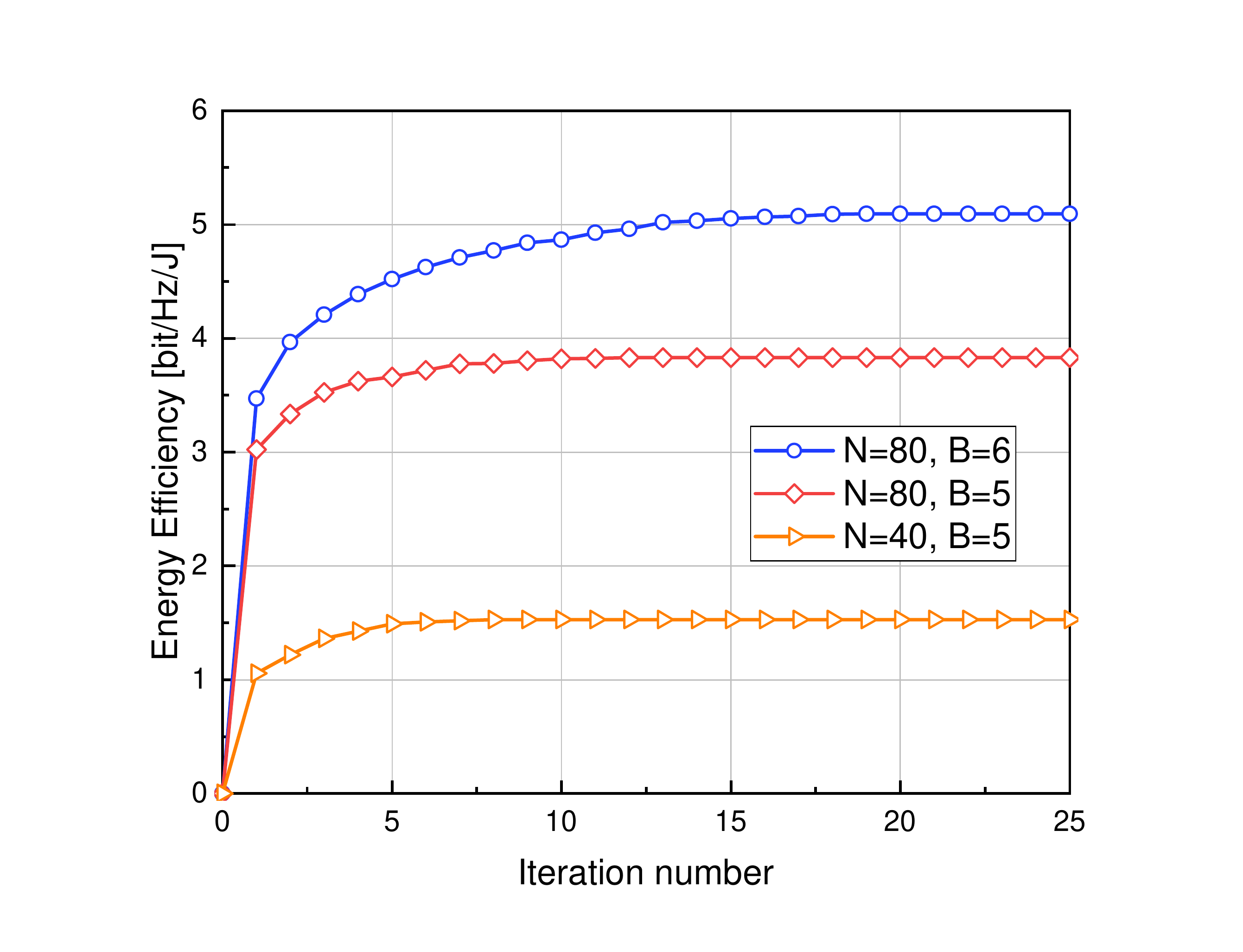}
	\vspace{-1.2em}
	\caption{Convergence performance of proposed algorithm with $K=12$ and $N_{RF}=20$.}
	\label{EE_Iter}
	\vspace{-1.5em}
\end{figure}

\begin{figure}
	[t]
	\vspace{-1em}
	\centering
	\includegraphics[width=.9\columnwidth,height=6cm]{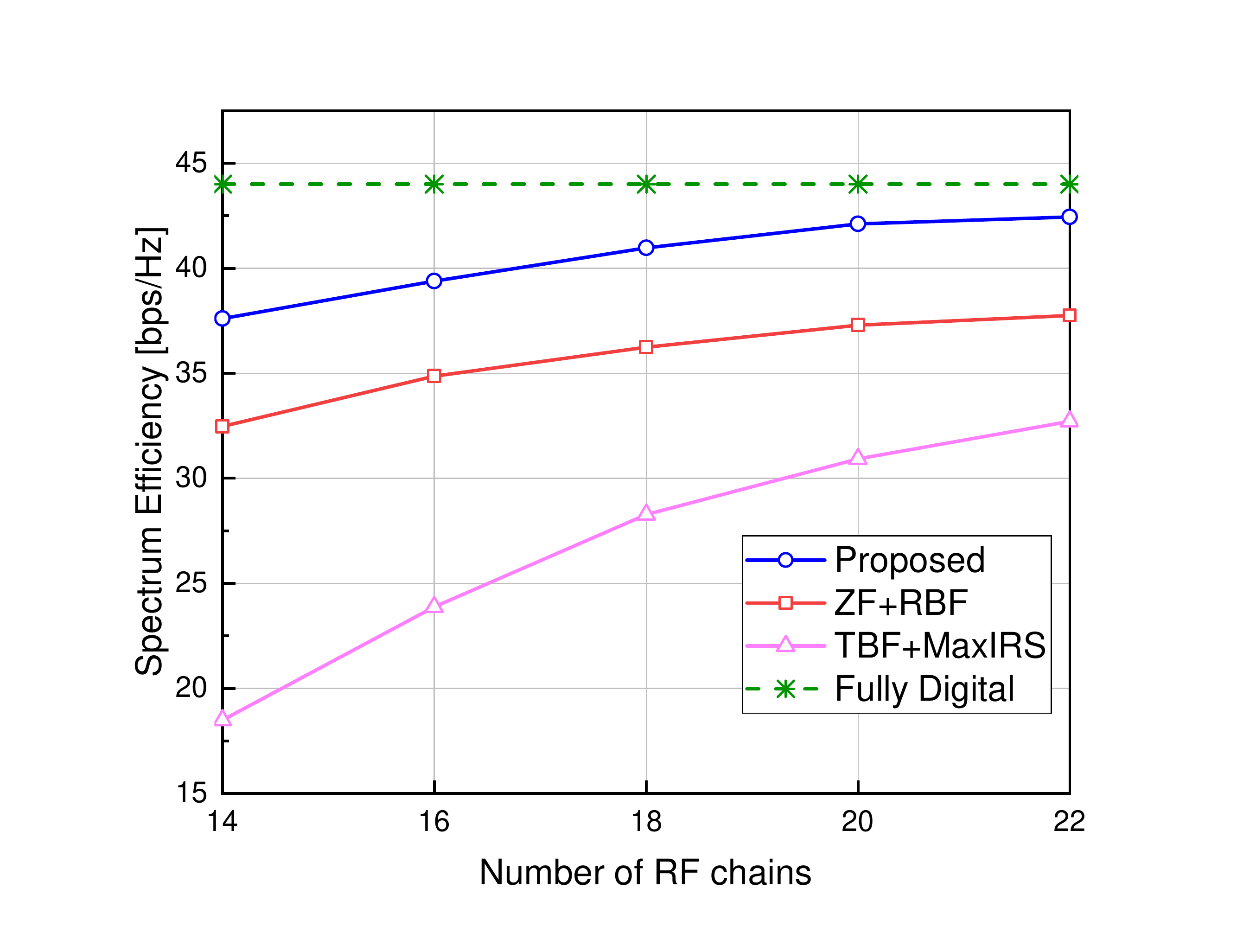}
	\vspace{-1.2em}
	\caption{SE versus $N_{RF}$ with $K=12$ and $N=80$.}
	\label{SE_RF}
	\vspace{-1.5em}
\end{figure}

In the simulations, we compare our proposed algorithm with other three schemes : (1) ``ZF + RBF", where ZF is utilized at BS and the reflect beamforming at IRSs is the same as our proposed algorithm; (2) ``TBF + MaxIRS", our proposed transmit beamforming is exploited at BS and ``MaxIRS" scheme is utilized at IRSs, \emph{i.e.}, ``MaxIRS" selects the phase shifts from the feasible set which maximize $\frac{1}{K}\sum_{k=1}^{K}\left|\mathbf{H}_k^H\boldsymbol{\omega}_k\right|$; (3) ``Fully Digital", where each antenna at BS is connected to one RF chain.

\begin{figure}
	[t]
	\centering
	\includegraphics[width=.9\columnwidth,height=6cm]{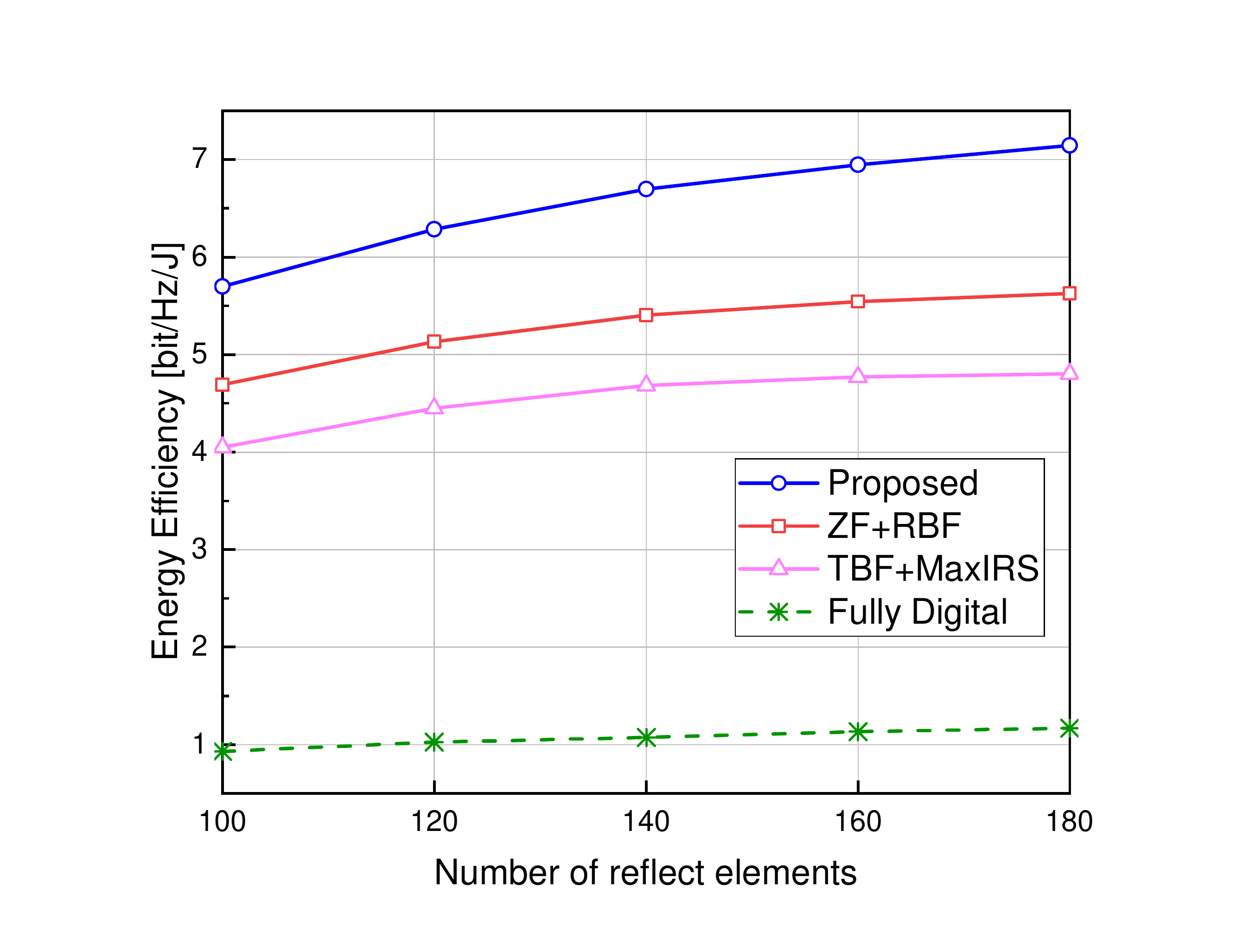}
	\vspace{-1.2em}
	\caption{EE versus $N$ with $K=12$ and $N_{RF}=20$.}
	\label{EE_N}
	\vspace{-2em}
\end{figure}

\begin{figure}
	[t]
	\centering
	\includegraphics[width=.9\columnwidth,height=6cm]{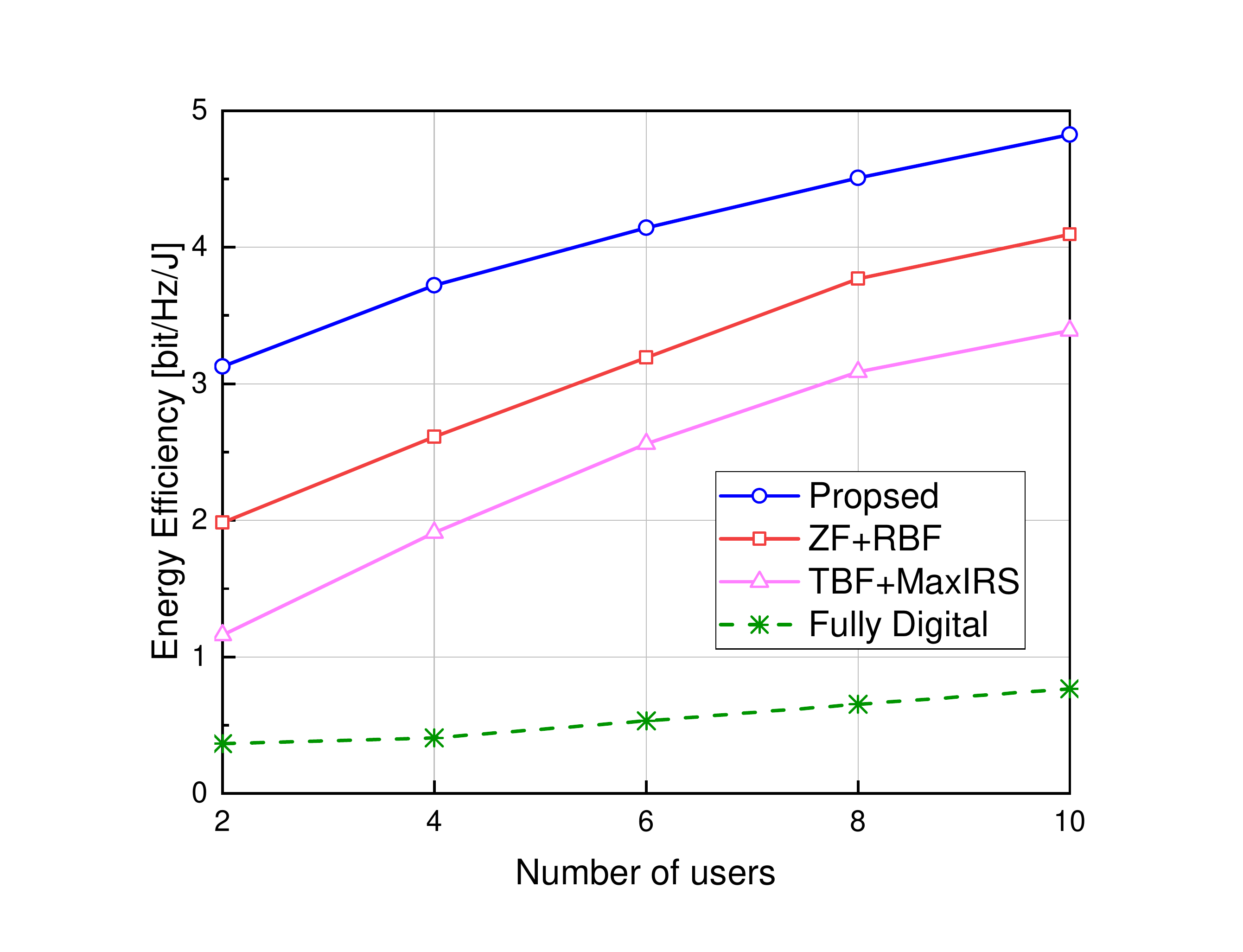}
	\vspace{-1.0em}
	\caption{EE versus $K$ with $N=80$ and $N_{RF}=20$.}
	\label{EE_K}
	\vspace{-1.5em}
\end{figure}

For simplicity, the following figures are plotted without multiplying the constant $W$. The convergence behavior of our proposed algorithm is presented in Fig. \ref{EE_Iter}. It can be seen that, the minimum $B$ and $N$ converges fastest but the upper bound is much lower than others. With increasing $B$ and $N$, our proposed algorithm converges slower but obtains significant improvement on EE.

In Fig. \ref{SE_RF}, we plot the spectrum efficiency (SE) against the number of RF chains at BS. As a tendency, the SE increases with the increase of RF chains number and the gap between our proposed algorithm and fully digital becomes smaller. This is because fully digital can be considered as a special case, \emph{i.e.}, $N_{RF}=M$. Meanwhile, our proposed algorithm outperforms the other two schemes, since ZF ignores the impact of noise and MaxIRS only maximizes the average received signal strength. 

Although fully digital achieves better SE, the EE decreases sharply due to the huge energy consumption of massive RF chains at BS as shown in Fig. \ref{EE_N} and Fig. \ref{EE_K}. The result in Fig. \ref{EE_N} shows that all these curves ascend since increasing reflect element number enables IRS to receive more signal energy from BS and reflect more effective signal to the users. Fig. \ref{EE_K} depicts EE in terms of user number. When $K$ increases moderately, the improvement of sum rate with individual QoS constraint has a greater impact on EE compared with the power promotion. As a result, the four curves increase with $K$ increasing, while our proposed algorithm provides the highest EE promotion due to the effective EE optimization algorithm.


\section{conclusion}
In this paper, we design an IRSs-enhanced MU mmWave lens system to overcome the drawbacks of mmWave, where the RF chains cost is efficiently reduced by lens antenna array, and IRSs are utilized to mitigate the blocking effect of mmWave and enhance the system. We then formulate an EE maximization problem with constraints of maximum data transmission power, individual QoS and limited RF chains number. Due to the non-convexity, this problem is solved by our proposed algorithm through the alternating optimization technique. Specifically, we optimize the transmit beamforming at BS by the SCA method and the reflect beamforming at IRSs is handled through the quadratic transform method. Simulation results validate that the joint optimization can achieve significant EE promotion compared with other schemes under various scenarios. Meanwhile, the convergence performance of our proposed algorithm has been provided and analyzed. In conclusion, our proposed algorithm can be utilized to guide the practical system deployment, to achieve higher system EE.


\bibliographystyle{IEEEtran}
\vspace{-0.5em}
\bibliography{reference}

\end{document}